\documentclass[twocolumn,english,aps,PRB,groupedaddress,floats]{revtex4}
\usepackage[T1]{fontenc}
\usepackage[latin9]{inputenc}
\usepackage{graphicx}



\usepackage{babel}

\begin{document}

\title{Band-edge Exciton States in a Single-walled Carbon Nanotube Revealed by Magneto-optical Spectroscopy in Ultra-high Magnetic Field}

\author{Weihang Zhou,$^{1}$ Tatsuya Sasaki,$^{1,4}$ Daisuke Nakamura,$^{1}$ Huaping Liu,$^{2,3}$ Hiromichi Kataura,$^{2,3}$ and Shojiro Takeyama$^{1}$}

\email{takeyama@issp.u-tokyo.ac.jp}

\affiliation{$^{1}$The Institute for Solid State Physics, The University of Tokyo, 5-1-5, Kashiwanoha, Kashiwa, Chiba 277-8581, Japan}

\affiliation{$^{2}$Nanosystem Research Institute, National Institute of Advanced Industrial Science and Technology, Tsukuba, Ibaraki 305-8562, Japan}

\affiliation{$^3$Japan Science and Technology Agency, CREST, Kawaguchi, Saitama 330-0012, Japan}

\affiliation{$^4$Department of Applied Physics, The University of Tokyo, Hongo 113-8656, Japan}

\date{\today}

\begin{abstract}
We report high field magneto-optical study on the first and second sub-band transitions of single-chirality single-walled carbon nanotubes. The ordering and relative energy splitting between bright and dark excitonic states were found to be inversed between the first and second subbands. We verified that the zero-momentum dark singlet exciton lies below the bright exciton for the first sub-band transitions, while for the second sub-band transitions, it was found to have higher energy than the bright excitonic state. Effect of this peculiar excitonic structure was found to manifest itself in distinctive Aharonov-Bohm splitting in ultra-high magnetic fields up to 190 T.
\end{abstract}


\maketitle

Single-walled carbon nanotubes (SWNTs) have attracted much attention since their discovery in the early 1990s \cite{Iijima1991, Charlier2007, Barros2006, Dresselhaus2005, Bachilo2002}. Their unique one-dimensional tubular structures of carbon network cause strong Coulomb interactions between carriers and are expected to lead to superb optical properties. However, it was found, surprisingly, that the photoluminescence (PL) quantum yield, i.e., the probability of re-emission after photon absorption, was extremely low (10$^{-4}$ to 10$^{-3}$) \cite{Connell2002, Wang2004}. The existence of optically inactive (dark) excitonic state below the first optically active (bright) excitonic state was generally ascribed to be the reason of this low PL quantum yield \cite{Shaver2007, Perebeinos2004, Perebeinos2005, Spataru2005}. However, exceptions were also predicted. Recent theoretical study shows that the ordering and relative energy splitting of SWNT excitonic states depend sensitively on the details of the intra-valley and inter-valley Coulomb interactions \cite{Ando2006}. As an important factor for SWNT optical properties, revealing this peculiar excitonic characteristics experimentally is not only interesting from the point of view of fundamental physics, but also key to the application of SWNTs in future opto-electronics.

On the other hand, it is now well known that an external magnetic field applied along SWNT axis changes their excitonic structures drastically. The magnetic flux threading the tubes breaks the time reversal symmetry and lifts the degeneracy between the K and K$^{'}$ states, eventually producing two independent bright excitonic states corresponding to the KK and K$^{'}$K$^{'}$ excitons for sufficiently large magnetic flux \cite{Ando2006, Zaric2004, Ando2004, Zaric2006, Mortimer2007, Takeyama2011}. The magnetic brightening of SWNT dark states is indeed an ideal test-bed for the study of Aharonov-Bohm (AB) effect in realistic systems. Moreover, what should be noted is that, it provides an efficient way to reveal the unknown relative ordering and energy splitting between bright and dark excitonic states of SWNTs.

In this work, we prepared stretch-aligned single-chirality (6,5) SWNTs and performed high-field magneto-absorption studies for both first (E$_{11}$) and second (E$_{22}$) sub-band transitions, with the aim of clarifying the mechanism of the relative ordering between dark and bright excitonic states. As an ideal realistic model for the study of AB effect, extensive spectroscopic work has been done for the E$_{11}$ transition in external magnetic fields \cite{Zaric2004, Takeyama2011, Srivastava2008, Matsunaga2008}. However, it is noteworthy that the higher energy sub-band transitions, especially their responses to the external magnetic field, remain mysterious so far. As essential parts for SWNT optical processes, it is of immense practical importance to reveal the excitonic structures, as well as their magnetic responses, for these higher energy sub-band transitions. Meanwhile, unlike those SWNT mixtures conventionally used so far in previous studies, the samples used in this work essentially contain only one specific chirality (6,5). The exceptionally high purification excludes disturbance from other species, yielding explicit optical spectra with which we could clearly follow spectral changes in magnetic fields. This is particularly important for the study of higher energy sub-bands, as their broadened absorption peaks could easily obscure weak signals. On the other hand, the AB splitting of SWNT excitonic states is proportional to the magnetic flux along tube axis. Therefore, successful preparation of stretch-aligned SWNTs significantly enhances the effective magnetic field along SWNT axis, reducing field convolution effect and thus facilitating reliable data analyses. To the best of our knowledge, this is the first systematic and precise experimental study for the high magnetic field effects on the higher energy sub-band transitions of SWNTs.

\begin{figure}[t!]
\includegraphics[width=8.5cm]{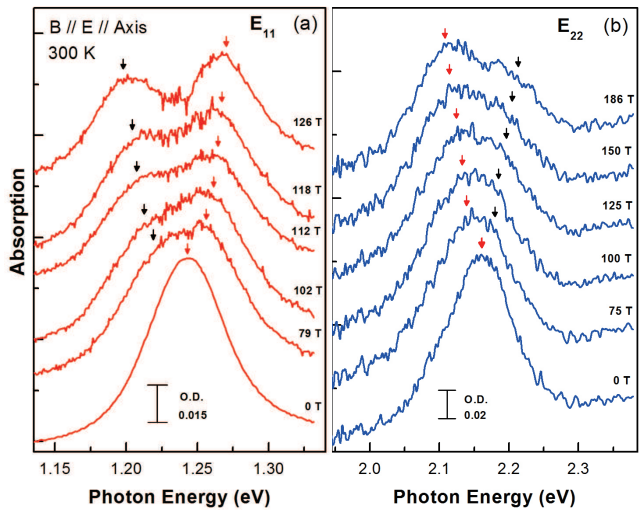}
\caption{(color online) Typical magneto-absorption spectra of the stretch-aligned (6,5) SWNTs for (a) E$_{11}$ and (b) E$_{22}$ transitions. The red (black) arrow denotes the peak position of the bright (dark) exciton at each field. Measurements were taken at room temperature under Voigt geometry, with light polarization and magnetic field B parallel to the stretching direction of the SWNTs/PVA film.} \label{LDAorbital}
\end{figure}
The samples we used are highly-selected single-chirality (6,5) SWNTs, which were separated from a high-pressure carbon monoxide (HiPco)-grown mixture by the single-surfactant multicolumn gel chromatography method \cite{Liu2011}. High degree of tube alignment was achieved by embedding SWNTs into polyvinyl alcohol (PVA) film, followed by a stretching process of the SWNTs/PVA film with stretching ratio of $\sim$ 5. During magneto-absorption measurements, a Xe-flash arc lamp was used as the light source. The ultra-high magnetic field up to 190 T was generated by the single-turn coil technique with typical pulse width of $\sim$ 7  $\mu$s \cite{Miura2003}. For E$_{11}$ measurement, an InGaAs diode array (XEVA-640, Xenics) was used as the detector. Spectra were taken on top of the pulsed magnetic field synchronously. Exposure time of the detector was chosen to be 1 $\mu$s to keep field variation within 4$\%$ during the gate opening time. While for E$_{22}$ measurements, a streak camera (C4187-25S, Hamamatsu Photonics) was used and thus magneto-absorption spectrum could be monitored continuously for the whole duration time of the pulsed field \cite{Miura1998}. All the spectra were taken under the Voigt geometry at room temperature.

\begin{figure}[t!]
\includegraphics[width=8.5cm]{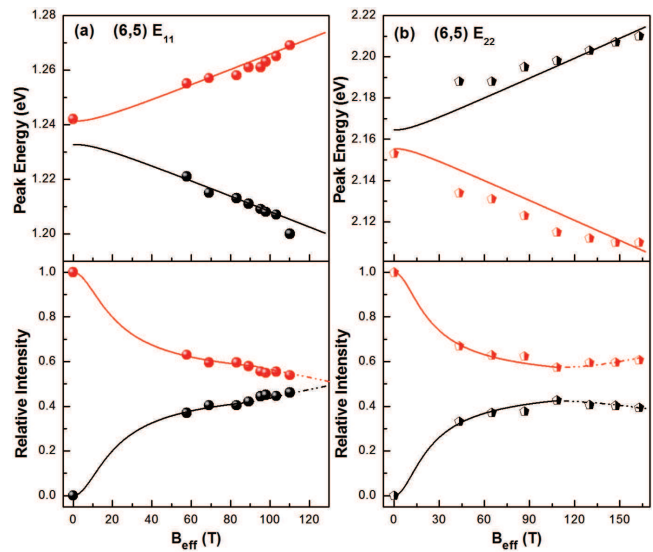}
\caption{(color online) (a) Energy and intensity of the bright (red filled circle) and dark (black filled circle) excitons as a function of the effective magnetic fields for E$_{11}$ transition. (b) Energy and intensity of the bright (red pentagon) and dark (black pentagon) excitons as a function of the effective magnetic fields for E$_{22}$ transition. Red and black solid lines are best-fit curves for the evolution of bright and dark excitons, respectively. The effective magnetic field $B_{\text{eff}}=B\cos\theta$.} \label{LDAorbital}
\end{figure}
Figure 1(a) shows the typical magneto-absorption spectra of the stretch-aligned (6,5) SWNTs for the E$_{11}$ transition. Without magnetic field, an absolute single peak, corresponding to the E$_{11}$ bright exciton, can be seen clearly without almost any background absorption. However, upon the application of an external magnetic field, obvious peak splitting takes place. For magnetic fields higher than 110 T, two peaks with comparable intensity could be identified. Moreover, from the evolutions of the two peaks, it is clearly demonstrated that the splitting is due to the emergence of a new peak on the low energy side of the bright exciton peak. It is now generally known that an external magnetic field mixes the wave functions of the bright and dark excitons with zero angular momentum. The mixing of wave functions redistributes the oscillator spectral weight between the two excitons and leads to the brightening of the dark excitonic state, which is a manifestation of the AB effect \cite{Ando2006, Zaric2004, Ando2004, Zaric2006, Mortimer2007, Takeyama2011}. Thus, the emergence of a new peak on the low energy side of the bright exciton peak shows, unambiguously, that the zero-momentum dark excitonic state lies below the E$_{11}$ bright excitonic state.

The AB splitting of SWNT excitonic states is proportional to the effective magnetic field applied parallel to the tube axis. For ensembles of SWNTs, effects of ensemble average could be accounted for by introducing the nematic order parameter, whose definition is $S = (3 < \cos^{2}\theta  > - 1) / 2$, with $\theta$ being the average angle between SWNT axis and the direction of external magnetic field \cite{Takeyama2011}. Exact value of the average angle $\theta$ could be deduced by correlating $S$ to the optical anisotropy $A = (\alpha_{\parallel}-\alpha_{\perp})/(\alpha_{\parallel}+2\alpha_{\perp})$, where $\alpha_{\parallel}$ and $\alpha_{\perp}$ denote the absorption intensities for light polarized parallel and perpendicular to the SWNT axis, respectively. Our measurement of optical anisotropy gives $S \approx 0.66$, yielding an average angle $\theta \approx 29^{\circ}$.

To extract the evolutions of the bright and dark excitons in magnetic fields precisely, we deconvoluted the magneto-absorption spectra with two Gaussian waveforms that give a very good fitting to our experimental data. Results of deconvolutions were shown in Figure 2(a) as a function of the effective magnetic field ($B_{\text{eff}}=B\cos\theta$) for the bright (red circle) and dark
\begin{table}[!h]
\tabcolsep 0pt \caption{Fitting parameters for the AB splittings of the first and second sub-band transitions.} \vspace*{-12pt}
\begin{center}
\def\temptablewidth{0.5\textwidth}
{\rule{\temptablewidth}{1pt}}
\begin{tabular*}{\temptablewidth}{@{\extracolsep{\fill}}ccccccc}

           &           & $\mu_{ex}$ & $\mu_{th}$ & Ratio               & $\Delta_{bd}$    & Lower  \\
           & Chirality & (meV/T)    & (meV/T)    & $\mu_{ex}/\mu_{th}$ & (meV)            & state \\ \hline
 E$_{11}$  & (6,5)     & 0.57       &  0.74      & 0.77                & $8.5\pm1.5$      & dark  \\
 E$_{22}$  & (6,5)     & 0.65       &  0.74      & 0.88                & $-(9.0\pm2.5)$   & bright  \\

\end{tabular*}
{\rule{\temptablewidth}{1pt}}
\end{center}
\end{table}
(black circle) excitons, respectively. According to previous studies, the AB splitting of SWNT excitonic states can be described well by the following formulae \cite{Shaver2007, Takeyama2011, Matsunaga2008}:
\begin{eqnarray}
&\varepsilon_{\beta,\delta}(B) = E_{g} \pm \frac{\sqrt{\Delta^2_{bd}+\Delta^2_{AB}(B)}}{2} \quad \text{(peak energy)}\\
&I_{\beta,\delta}(B) = \frac{1}{2} \pm \frac{1}{2} \frac{\Delta_{bd}}{\sqrt{\Delta^2_{bd}+\Delta^2_{AB}(B)}} \quad \text{(peak intensity)}
\end{eqnarray}
where $\Delta_{bd}$ represents the zero-field splitting between the bright ($\beta$) and dark ($\delta$) excitons. $\Delta_{AB} = \mu B \cos \theta$ is the so-called AB splitting. Here $E_{g}=(E_{b}+E_{d})/2$, is the center of the bright and dark exciton level in the absence of magnetic field. Fitting results using the above formulae were shown as solid lines in Figure 2(a), showing the excellent agreement between the experimental data and the fitted curves. Parameters used in this fitting are $\Delta_{bd} = 8.5 \pm 1.5$ meV and $\mu \approx 0.57$ meV/T, which are selected so as to fit the evolutions of peak energy as well as peak intensity.

Having extracted the details for the magnetic field evolutions of the E$_{11}$ transition, we now turn our attention to the higher energy E$_{22}$ transitions, with the aim of clarifying the mechanism of the relative ordering between bright and dark excitonic states. Typical magneto-absorption spectra up to 186 T were shown in Figure 1(b). A single peak, corresponding to the E$_{22}$ bright exciton, was observed clearly in the absence of magnetic field. Again, as an external magnetic field is applied, peak splitting appears due to the AB effect. However, what should be noted here is that, the new peak emerges on the high-energy side of the E$_{22}$ bright exciton peak, as indicated by the black arrow in Figure 1(b). Considering the mechanism of AB splitting for SWNTs, the distinctive splitting of E$_{22}$ demonstrates unambiguously that the zero-momentum dark excitonic state lies above the bright excitonic state for the E$_{22}$ manifold, in strong contrast with the first sub-band E$_{11}$ transition.

Again, to extract the details of the magnetic field evolutions, we deconvoluted the magneto-absorption spectra using two Gaussian waveforms. Results of the Gaussian fittings were shown in Figure 2(b) as red and black pentagons for the bright and dark excitonic components, respectively. The evolutions of the bright and dark excitons were fitted using the same formulae as the E$_{11}$ transition, as shown by the solid lines in Figure 2(b). The fitting gives $\Delta_{bd} = - (9.0 \pm 2.5)$ meV and $\mu \approx 0.65$ meV/T.

Details of the parameters obtained from the AB splittings of the first and second sub-band transitions were summarized in Table 1. It is interesting to note that $\mu$, the coefficient characterizing the AB effect of SWNTs, was obviously smaller than its theoretical value $\mu_{th}$, where $\mu_{th} = 3 (\pi e d^{2} / 2h)E_{g}$ ($d$: diameter). For the first sub-band transitions, $\mu_{ex}$ is $77\%$ of $\mu_{th}$, while for the second sub-band transitions, $\mu_{ex}$ is $88\%$ of $\mu_{th}$. These results are in qualitative agreement with previous reports using SWNTs with chiral mixtures \cite{Takeyama2011}. Thus, the overestimation of $\mu$ is probably due to the oversimplification of the present model. Field-induced effects, such as the magnetic field-dependent exciton binding energy, should be included to give more accurate predictions.

\begin{figure}[t!]
\includegraphics[width=8.5cm]{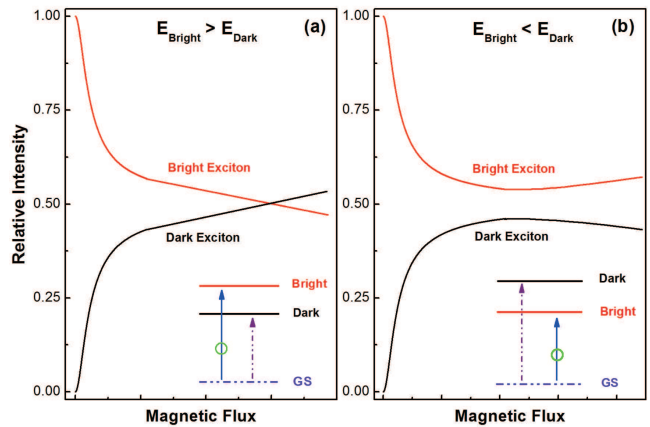}
\caption{(color online) Schematics for the intensity evolutions of the bright and dark excitons in external magnetic fields. (a) Bright excitonic state lies above the dark excitonic state; (b) Bright excitonic state lies below the dark excitonic state. The green circles in the insets represent the optically allowed transitions.} \label{LDAorbital}
\end{figure}
Evolutions of the bright and dark excitons were described well by equations (1) and (2) for both first and second sub-band transitions, despite their difference in the relative ordering of the bright and dark excitonic components. The completely different ordering between the bright and dark excitonic states, however, results in distinctive intensity evolutions in sufficiently high magnetic fields. The external magnetic field lifts the degeneracy between the K and K$^{'}$ states, producing two bright excitons corresponding to the independent KK and K$^{'}$K$^{'}$ excitons for sufficiently strong magnetic fields. In the case that the dark exciton lies below the bright exciton, the dark excitonic state is shifted to lower energy, while the bright excitonic state is shifted toward higher energy,  due to the AB flux. Therefore, the optical intensity of the dark exciton would be eventually larger than that of the bright exciton, as a result of the decrease of the dark exciton energy in external magnetic fields \cite{Ando2006, Ando2004, Ajiki1993, Ando2005}. A schematic diagram, similar to that in Ref [12], was shown in Figure 3(a), as a guide for eyes. However, in the case that the dark exciton is higher in energy, the dark excitonic state would be shifted to higher energy while the bright excitonic state moves in the lower energy direction. The dark excitonic state, which is brightened by the external magnetic field, gains oscillator strength gradually, at the expense of the bright excitonic state. After it reaches some critical intensity at certain magnetic field, its intensity decreases again since it lies above the bright exciton. Schematic for this peculiar intensity evolution was shown in Figure 3(b). Comparing with the experimental data shown in the lower panels of Figure 2, it is demonstrated, unambiguously, that the intensity evolutions of the first and second sub-band transitions mimic those shown in Figure 3 (a) and (b), respectively. This peculiar intensity evolution leads us to confirm again, that the dark exciton lies above the bright exciton for the second sub-band transitions, while for the first sub-band transitions, it is opposite. Furthermore, it should be noted that, equation (2) describes the intensity evolution well only at relatively low magnetic fields due to its simplication and the complicated intensity evolutions of the bright and dark excitons at high magnetic fields. On the other hand, the energies of the bright and dark excitons show relatively simple monotonic evolvements even in ultra-high magnetic fields. Thus, equation (1) still gives a good fitting to the experimental data at high magnetic fields. For safety, the fittings using equation (2) were carried out only when the effective magnetic field ($B_{\text{eff}} = B \cos \theta$) is below 90 T for E$_{11}$ and 110 T for E$_{22}$. For the first sub-band transitions, the critical field at which bright and dark excitons have the same oscillator strength was estimated to be $\sim$ 130 T by means of extrapolation, as shown by the dash-dotted lines in Figure 2(a). For the second sub-band transitions, it can be seen clearly from Figure 2(b) that the intensities of the bright and dark excitons are closest to each other at $\sim$ 110 T. Further increase of magnetic field causes larger difference of their intensities, in excellent agreement with theoretical predictions.

In summary, we performed ultra-high field magneto-optical spectroscopic studies on both first and second sub-band transitions of SWNTs with purely single (6,5) chirality. The AB splitting of SWNT excitons was observed for both first and second sub-band transitions. The ordering and relative energy splitting of the dark and bright excitonic states were investigated by means of the AB effect. From the evolutions of the excitonic AB splitting, we verified that the relative ordering between the bright and dark excitons was highly sub-band index dependent. The zero-momentum dark excitonic state lies below the bright exciton for the first sub-band transitions, while for the second sub-band transitions, it was found to have higher energy than the bright excitonic state. The relative ordering of bright and dark excitons is critical to the optical properties of SWNTs, our work thus would serve as a guide for future studies as well as potential applications in opto-electronics.

\acknowledgements{We thank Prof. H. Suzuura at Hokkaido University for fruitful discussions. Technical support of sample preparation from Prof. S. Maruyama and Mr. S. Harish at the University of Tokyo is gracefully acknowledged.}

\end{document}